\documentclass[runningheads]{llncs}

\usepackage{accv}

\usepackage{accvabbrv}

\usepackage{graphicx}
\usepackage{booktabs}
\usepackage{subcaption}

\usepackage[accsupp]{axessibility}  % Improves PDF readability for those with disabilities.

\usepackage{hyperref}

\usepackage{orcidlink}
\usepackage{enumitem}

\begin{document}

\title{Feature Estimation of Global Language Processing in EEG Using
Attention Maps} 

\titlerunning{Abbreviated paper title}

\author{
Dai Shimizu\inst{1}\orcidlink{0009-0001-4209-6608} \and
Ko Watanabe\inst{2,3}\orcidlink{0000-0003-0252-1785} \and
Andreas Dengel\inst{2, 3}\orcidlink{0000-0002-6100-8255}
}

\authorrunning{D.~Shimizu et al.}

\institute{
Tokyo Institute of Technology, 2-12-1 Ookayama, 152-8550, Tokyo, Japan\\ \email{shimizu.d.ab@m.titech.ac.jp} \and
RPTU Kaiserslautern-Landau, Erwin-Schrödinger-Straße 52, 67663, Kaiserslautern, Germany \and
DFKI GmbH, Trippstadter Str. 122, 67663, Kaiserslautern, Germany \\
\email{\{first.last\}@dfki.de}}

\maketitle

% \author{Dai Shimizu}
% \email{daistar1202@gmail.com}
% \orcid{0009-0001-4209-6608}
% \affiliation{%
%   \institution{Tokyo Institute of Technology}
%   \city{Tokyo}
%   \country{Japan}
% }

% \author{Ko Watanabe}
% \orcid{0000-0003-0252-1785}
% \email{ko.watanabe@dfki.de}
% \affiliation{%
%   \institution{RPTU Kaiserslautern-Landau {\&} DFKI GmbH}
%   \city{Kaiserslautern}
%   \country{Germany}
% }

% \author{Andreas Dengel}
% \orcid{0000-0002-6100-8255}
% \email{andreas.dengel@dfki.de}
% \affiliation{%
%   \institution{RPTU Kaiserslautern-Landau {\&} DFKI GmbH}
%   \city{Kaiserslautern}
%   \country{Germany}
% }

\begin{abstract}
  Understanding the correlation between EEG features and cognitive tasks is crucial for elucidating brain function. Brain activity synchronizes during speaking and listening tasks. However, it is challenging to estimate task-dependent brain activity characteristics with methods with low spatial resolution but high temporal resolution, such as EEG, rather than methods with high spatial resolution, like fMRI. This study introduces a novel approach to EEG feature estimation that utilizes the weights of deep learning models to explore this association. We demonstrate that attention maps generated from Vision Transformers and EEGNet effectively identify features that align with findings from prior studies. EEGNet emerged as the most accurate model regarding subject independence and the classification of Listening and Speaking tasks. The application of Mel-Spectrogram with ViTs enhances the resolution of temporal and frequency-related EEG characteristics. Our findings reveal that the characteristics discerned through attention maps vary significantly based on the input data, allowing for tailored feature extraction from EEG signals. By estimating features, our study reinforces known attributes and predicts new ones, potentially offering fresh perspectives in utilizing EEG for medical purposes, such as early disease detection. These techniques will make substantial contributions to cognitive neuroscience.
\end{abstract}

\section{Introduction}
In recent years, electroencephalography (EEG) has emerged as a critical instrument for real-time monitoring of brain activity, owing to its superior temporal resolution and non-invasive nature. However, analyzing EEG data remains a complex task due to inter-individual variability and the subtlety of neural signals. 

Traditional signal processing techniques often find it challenging to handle the complexity of EEG data and effectively extract task-specific features.
Deep learning models, especially those equipped with attention mechanisms like the Transformer~\cite{vaswani2017attention} and Vision Transformer(ViT)~\cite{dosovitskiy2020image}, have proven to be powerful tools in various domains, including image identification, natural language processing, and complex biological signal analysis. These models offer significant improvements over conventional methods by highlighting relevant features in extensive datasets.
The computation of attention mechanism weights, as in Gradient-weighted Class Activation Mapping (Grad-CAM)~\cite{selvaraju2017grad} or Vision Transformer for Attention Map, can identify areas of interest for classification results. Interestingly, these areas can also be computed from the classification results themselves.

This study's novelty lies in using such neural network models, specifically the Vision Transformer, to estimate features from the attention map. These features are not specific to EEG but are rough, task-dependent features. This approach allows us to implement subject-independent analysis and use the entire language-related area~\cite{binder1997human,hollenstein2021decoding} of the brain for training dataset creation.
Interestingly, during speaking and listening activities, brain activity becomes similar~\cite{li2023eeg,kuhlen2012content,perez2017brain}. In this context, we aim to estimate the features of speaking and listening using the weights of the neural network model. By leveraging the internal weights of these models to compute attention maps, this study aims to uncover subtle EEG signal patterns indicative of specific brain functions.

The contributions of this study are twofold:

    % \begin{itemize}
    
    %     \item[C1]\label{con:c1} The utility of estimating EEG features using ViT
    %     \item[C2]\label{con:c2} A comprehensive evaluation of auditory information in EEG using ViT
    
    % \end{itemize}

\begin{enumerate}[label=C\arabic*, leftmargin=*, itemsep=1em]
    \item \label{con:c1} The utility of estimating EEG features using ViT specifically focusing on EEG-based language processing
    \item \label{con:c2} A comprehensive evaluation of auditory information with participant independent in EEG using ViT
\end{enumerate}

These findings will revolutionize EEG data interpretation, enhancing diagnostic capabilities and personalizing neurotherapeutic approaches. This work is expected to make significant contributions to the fields of neuroimaging and cognitive neuroscience.

% The paper is structured as follows: a detailed methodology section is followed by the presentation and discussion of results. The paper concludes with insights drawn from the findings and suggestions for future research. This work underscores the potential of neural networks like Vision Transformer in EEG analysis, opening up new avenues for understanding and interpreting brain activity

\section{Related Work}
Deep learning techniques have significantly revolutionized the field of EEG analysis. This section first discusses the classification of EEG signals using various methods such as Power Spectral Density (PSD)~\cite{al2014methods}, EEGNet~\cite{lawhern2018eegnet}, and other CNN models~\cite{craik2019deep}. It then delves into applying Grad-CAM with CNNs and attention maps with Vision Transformers for feature extraction and interpretation in EEG signals.

In the realm of EEG signal classification, several techniques have been employed. PSD and other methods related to EEG frequency have been utilized in the context of emotion recognition, where these were extracted from EEG recorded during a listening task, revealing certain relevant frequency bands~\cite{alsolamy2016emotion,mihajlovic2019eeg,park2011seizure}. CNN-based architectures, particularly EEGNet~\cite{lawhern2018eegnet}, have been tailored for EEG signal processing, enabling automatic feature extraction and classification across various EEG analysis applications~\cite{sun2023cross}. Other CNN models using Net structures have also been widely adopted for EEG classification~\cite{liu2021tacnet}.

Grad-CAM~\cite{selvaraju2017grad} is a technique that enhances interpret-ability in models based on CNNs~\cite{watanabe2023engauge}. It highlights the critical regions within the input that influence the classification outcomes, thereby making the decision-making processes of CNNs transparent and comprehensible. This method has been instrumental in elucidating how CNNs prioritize different regions in an input image or signal during classification tasks.
The combination of EEGNet with Grad-CAM has been used to select the most suitable electrodes' channel~\cite{li2020eeg}. Moreover, the Grad-CAM technique in EEGNet was used to determine which brain area was involved in intention~\cite{leong2023ventral,orima2023spatiotemporal}.

ViT~\cite{dosovitskiy2020image} has been applied to EEG studies~\cite{arjun2021introducing,gong2023eeg,chen2023understanding}, marking a significant shift in image classification. ViTs use attention maps to illustrate how different image parts influence classification, providing insights into decision-making processes. In EEG studies, these maps reveal brain region activations during cognitive tasks, enriching our understanding of brain function.
Extensive studies have focused on delineating specific brain regions involved in auditory information processing~\cite{binder1997human}. Techniques, including EEG, have been pivotal in activating and studying various cerebral regions in response to auditory stimuli. Insights from this research are critical for comprehending auditory system functions and have profound implications for diagnosing and treating auditory-related disorders.
The application of advanced deep learning techniques such as Grad-CAM and Vision Transformers has markedly enriched EEG analysis. 

These methodologies boost the analytical capabilities and enhance the interpretability of EEG-based models, paving the way for significant neurological discoveries. Ongoing and future studies are expected to further harness the potential of these innovative techniques in complex EEG signal analysis.

\section{Methodology}
%\subsection{Experimental Paradigm}
In this study, our primary objective is to investigate whether the attention mechanisms in neural network models can capture the characteristics of brain waves depending on the task. Specifically, we aim to compare the brain waves recorded while listening to speech and while speaking the same speech heard. By examining these two conditions, we seek to identify broad differences in neural activity patterns associated with auditory perception and speech production.

To achieve this, we utilize several models: a pre-trained Vision Transformer (pre-trained ViT)~\cite{huggingface2023}, a customized Vision Transformer (Custom ViT)~\cite{dosovitskiy2020image}, EEGNet~\cite{lawhern2018eegnet}, and a Support Vector Machine (SVM)~\cite{schuldt2004recognizing}. These models classify data during the listening and speaking phases.
We aim to compute which aspects each neural network focuses on by analyzing the weights of these models. Subsequent sections will describe the detailed dataset types, data processing, and methods for creating attention maps for each model.

\subsection{EEG Data from OpenNEURO}

\begin{figure}[h]
\centering
\includegraphics[width=0.35\textwidth]{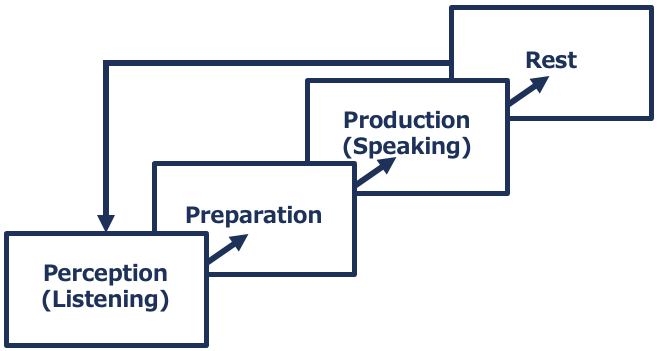}
\caption{Experimental protocol of the dataset. Subjects listen to and then repeat one of 30 randomly selected Spanish sentences, forming 30 perception-production pairs. Each sentence lasts approximately two seconds. Subjects perform between 360 and 420 trials, with each figure representing one trial.}
\label{fig:stimuli}
\end{figure}

This study uses a public dataset available from OpenNEURO~\cite{Valle} based on EEG recordings. The EEG data utilized in this study were obtained from Spanish participants. The dataset comprises 60 sessions, each recorded from 64 EEG channels and Electrocardiogram(ECG) and Electrooculogram (EOG) channels at a sample rate of 1000 Hz. This dataset was collected from 56 healthy participants.

The experimental paradigm presents participants with one of 30 different Spanish sentences, selected randomly for each trial. After listening to the sentence, participants are asked to repeat it aloud. Each sentence lasts approximately 2 seconds, and subjects perform between 360 and 420 trials. Figure~\ref{fig:stimuli} illustrates protocol ensures a comprehensive set of perception-production pairs for analysis.

\begin{figure}[t!]
\centering
\includegraphics[width=0.35\textwidth]{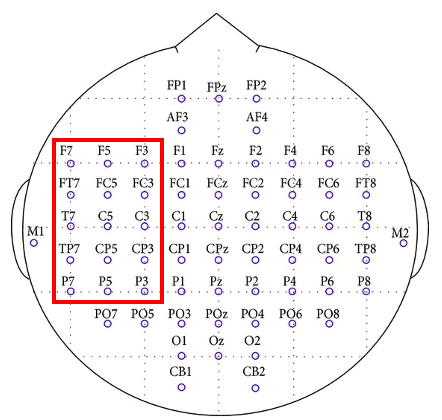}
\caption{Electrode placement following the 64-channel international 10–20 system. Electrodes framed in red were used.}
\label{fig:place}
\end{figure}

\subsection{Channel Selection for Classification}
This study investigates how broad auditory information is represented in brain waves. To achieve this, we utilized multiple channels from the EEG recordings to form a comprehensive dataset. The selection of specific channels is informed by existing literature, which indicates heightened activity in the left hemisphere, particularly the left temporal lobe, during auditory tasks~\cite{price2000anatomy,price2012review,de2017hierarchical}. Therefore, we focused on channels located in these regions to capture relevant neural activities.

As shown in Figure \ref{fig:place}, we specifically extracted EEG data from the following channels: 'F7', 'F5', 'F3', 'FT7', 'FC5', 'FC3', 'T7', 'C5', 'C3', 'TP7', 'CP5', 'CP3', ‘P7', 'P5', and 'P3'. These channels were chosen due to their significance in language processing tasks~\cite{binder1997human,hollenstein2021decoding}, which are the primary focus of our analysis. By isolating these particular channels, we aim to capture EEG features that are most indicative of the cognitive functions associated with language and auditory processing.

\subsection{Pre-processing and Data Processing}

\begin{table}[t!]
\centering
  \caption{Datasets: ALL indicates all participants, S indicates the number of trials applied, and C indicates the number of channels used.}
  \begin{tabular}{cccc}
    \toprule
    Model & Training Set & Validation Set & Test Set\\
    \midrule
    \texttt{Custom-ViT} & \((\text{ALL}-1) \times S \times C\) & \(\frac{1 \times S \times C}{2}\) & \(\frac{1 \times S \times C}{2}\)\\
    \texttt{Pretrained-ViT} & \((\text{ALL}-1) \times S \times C\) & \(\frac{1 \times S \times C}{2}\) & \(\frac{1 \times S \times C}{2}\)\\
    \texttt{EEGNet} & \((\text{ALL}-1) \times S\) & \(\frac{1 \times S}{2}\) & \(\frac{1 \times S}{2}\)\\
    \texttt{SVM} & \((\text{ALL}-1) \times S \times C\) & \(\frac{1 \times S \times C}{2}\) & \(\frac{1 \times S \times C}{2}\)\\
    \bottomrule
  \end{tabular}
  \label{tab:datasets}
\end{table}

We applied a band-pass filter to isolate frequencies from 1 to 40 Hz in EEG data, capturing the most relevant waves for cognitive and neural processes~\cite{bigdely2015prep}. We implemented artifact removal procedures for EOG and ECG signals to enhance the clarity of neural signal interpretation~\cite{bigdely2015prep}. We focused on language-related channels, such as Broca's and Wernicke's areas, to concentrate our analysis on the neural substrates of language function~\cite{binder1997human,hollenstein2021decoding,blank2002speech}. These preprocessing steps were uniformly applied across all computational models employed in our study to establish a consistent foundation for downstream analyses.

In this study, we leveraged Mel-spectrograms~\cite{stevens1937scale} to extract the spatio-temporal characteristics of EEG data for the training and evaluation of ViTs. The Mel-spectrogram transformation~\cite{daube2019simple} was selected due to its effectiveness in encapsulating the dynamic changes in EEG signal power across both time and frequency domains~\cite{abdul2022mel}, which is essential for our models to learn the intricate patterns associated with different cognitive states.
For the EEGNet architecture, which requires input in Channel and Time series, we downsampled the EEG data from 1000 Hz to 125 Hz. This preprocessing step was implemented to align with the Nyquist criterion~\cite{sun2011impedance}, ensuring the capture of all pertinent information below the 40 Hz frequency threshold, which encompasses the delta, theta, alpha, and beta wavebands known to be most relevant for brain-computer interface applications.
Finally, for the SVM classifier, we utilized PSD~\cite{alsolamy2016emotion,park2011seizure} estimates as the dataset to effectively reduce the dimensionality of the EEG signals. By transforming the data into the power frequency domain, we aim to highlight the most discriminative features for classification while simultaneously reducing computational complexity and enhancing model interpretability.

Additionally, a leave-one-subject-out (subject independence) approach was employed for all models to prevent the overlap of participant data during model training, ensuring that the training sets were participant-independent~\cite{albawi2018social}. Table~\ref{tab:datasets} shows the data split.

\subsection{Models Used for Data Analysis}
Our study adopted the Custom ViT to generate attention maps that span both temporal and frequency domains. The Custom ViT utilized the same structure described in the original paper~\cite{dosovitskiy2020image} and we also integrated a pre-trained ViT~\cite{huggingface2023}, utilizing its pre-trained weights and the same architecture to explore the interpretative capabilities of a network trained on extensive datasets with the following specific configurations:
    \begin{itemize}
    \item[-]pre-trained ViT: 12 layers, 12 heads, 16 patch size, 14x14 patches, 224x224 input, and 1024 MLP dimensions.
    \item[-]Custom ViT: 3 layers, 4 heads, 4 patch sizes, 8x8 patches, 32x32 input, and 256 MLP dimensions.
    \end{itemize}
For Custom ViT, all layers were considered for training, and for pre-trained ViT, only the final layer of Linear was considered for training.

For baseline comparisons, we employed EEGNet and SVM as standard models for EEG classification. EEGNet enables the extraction of temporal attention maps by applying Grad-CAM on the convolutional weights of the final layer.

During the model training phase, we employed a subject-wise cross-validation approach~\cite{kwon2019subject}. This involved using the data from a single subject as the validation and test set, while the remaining subjects' data constituted the training set. Such a strategy ensures that the model learns to generalize features of EEG data across different tasks and individuals, rather than overfitting to the characteristics of a single subject's data. This methodological choice is pivotal for developing robust EEG-based models that can reliably perform across diverse population samples, thereby enhancing the universality and applicability of the findings.

\subsection{Attention Maps}
This section describes the methodology employed to compute the attention maps for each model used in this study. Attention maps were utilized to explain the classification decisions made by the models, highlighting the features that contributed most significantly to their predictions.
For the Custom ViT and the pre-trained ViT, attention maps were derived from the weights of the final layer. This involved extracting the attention weights corresponding to the most significant parts of the input data, as identified by the model during classification. 

Specifically, the following steps were performed to compute the attention maps for the Vision Transformers:

\begin{enumerate}
\item The EEG data were transformed into mel-spectrograms, which were then fed into the ViTs.
\item The attention weights from the final layer were extracted, representing the importance of different time-frequency regions in the input data.
\item These weights were visualized to create the attention maps, illustrating the areas the model focused on during classification.
\end{enumerate}
For EEGNet, we used Grad-CAM to generate attention maps. Grad-CAM provides a visual explanation by highlighting the regions of the input that are most influential for the model's prediction. The following steps outline the process:
\begin{enumerate}
\item EEG data were input into the EEGNet model, which processes them through its convolutional layers.
\item Grad-CAM was applied to the convolutional weights of the final layer, identifying the most critical features for classification.
\item The resulting attention maps display the temporal regions and channels that contributed most to the model's decision.
\end{enumerate}

To ensure the reliability of the attention maps, we calculated them from the top 10 participants with the highest classification accuracy, denoted as '@10' in Table \ref{tab:acc}. This selection criterion helps focus on the dataset's most informative and consistent patterns. All attention maps were derived from the final layer of the models~\cite{woo2018cbam}. By utilizing attention maps from both ViTs and EEGNet, we aim to gain insights into the classification criteria used by each model, providing a clearer understanding of how neural network models interpret EEG data for task-related cognitive processes.

\subsection{Software and Tools}
All processing tasks, excluding data collection, were executed using Python. We utilized the PyTorch framework for deep learning algorithms, which is renowned for its flexibility and efficiency in building complex neural network architectures. EEG signal processing was conducted using the MNE library~\cite{gramfort2013meg}, which is specifically designed for advanced electrophysiological data analysis and provides robust tools for EEG data manipulation and visualization.

\section{Result}
\subsection{Accuracy of Classification}
As presented in Table~\ref{tab:acc}, EEGNet attained the highest classification accuracies among all models tested, recording values of 0.7248 for all participants and 0.8433 for the top 10 participants. While the ViTs, both Custom ViT and pre-trained ViT, did not achieve the highest overall accuracies, Custom ViT was notably the second most accurate model in the binary classification task: listening versus speaking across all participants.

\begin{table}[t!]
\centering
  \caption{Comparison of model accuracies for classification employed a leave-one-participant-out cross-validation approach, with one participant left out in each fold}
  \begin{tabular}{ccl}
    \toprule
    Model & Accuracy & Accuracy@10\\
    \midrule
    \texttt{SVM}& 0.5884 & 0.7048\\
    \texttt{Custom ViT} & 0.6153 & 0.6704 \\
    \texttt{pre-trained ViT}& 0.5633 & 0.6222\\
    \texttt{EEGNet}& \textbf{0.7248} & \textbf{0.8433}\\
    \bottomrule
  \end{tabular}
  \renewcommand{\arraystretch}{1}
  \vspace{0.2cm}
  \label{tab:acc}
\end{table}

\subsection{Attention Maps}

\begin{figure*}[htbp]
\captionsetup{}
    \centering
    \begin{subfigure}{.49\textwidth}
        \centering
        \includegraphics[width=\textwidth]{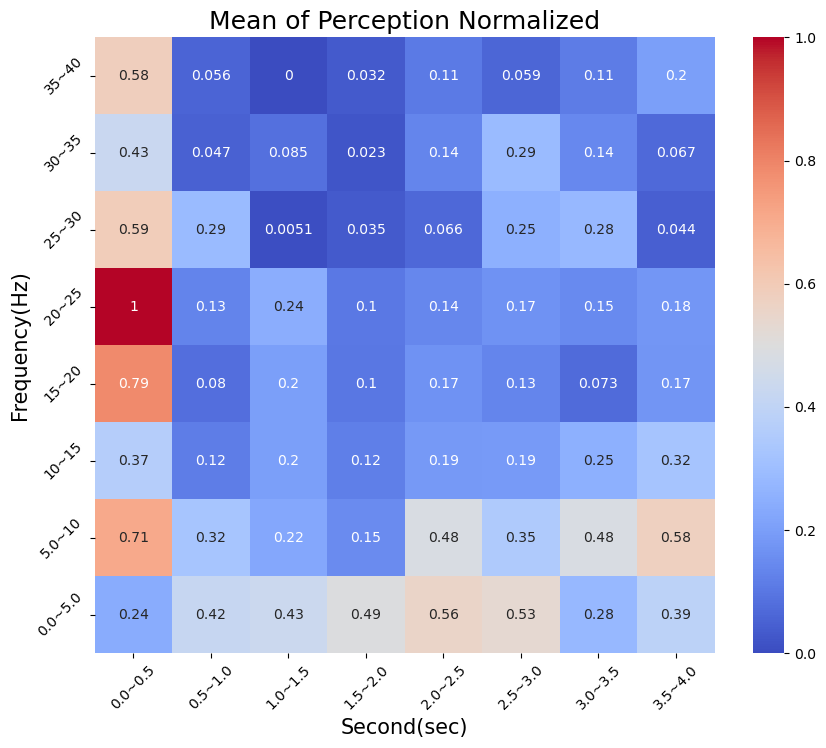}
        \caption{Attention map only for Perception tasks from Custom ViT}
        \label{fig:a}
    \end{subfigure}
    \hfill % Add separation between subfigures
    \begin{subfigure}{.49\textwidth}
        \centering
        \includegraphics[width=\textwidth]{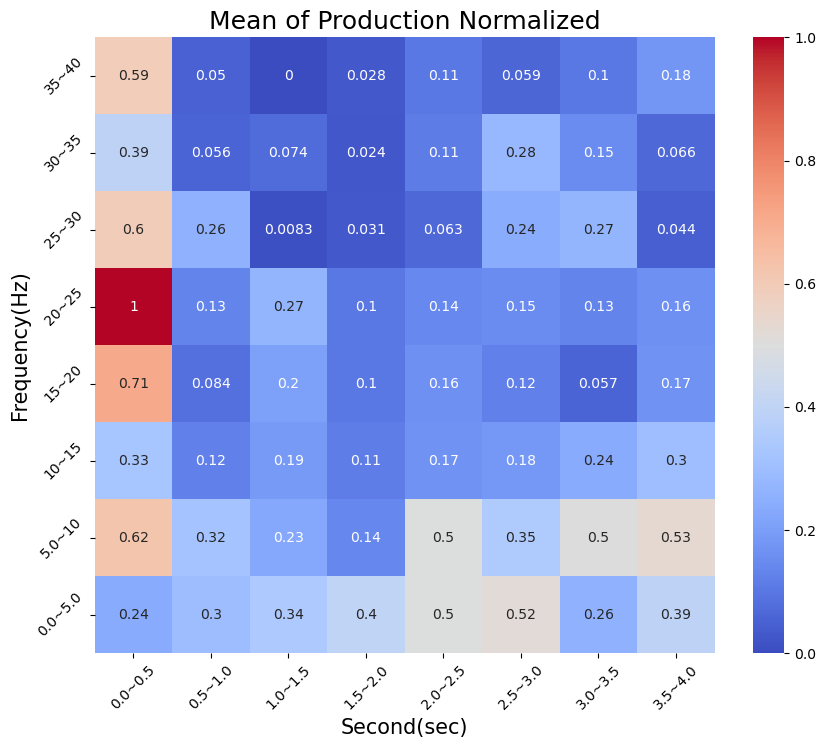}
        \caption{Attention map only for Production tasks from Custom ViT}
        \label{fig:b}
    \end{subfigure}
    \hfill % Add separation between subfigures
    \begin{subfigure}{.49\textwidth}
        \centering
        \includegraphics[width=\textwidth]{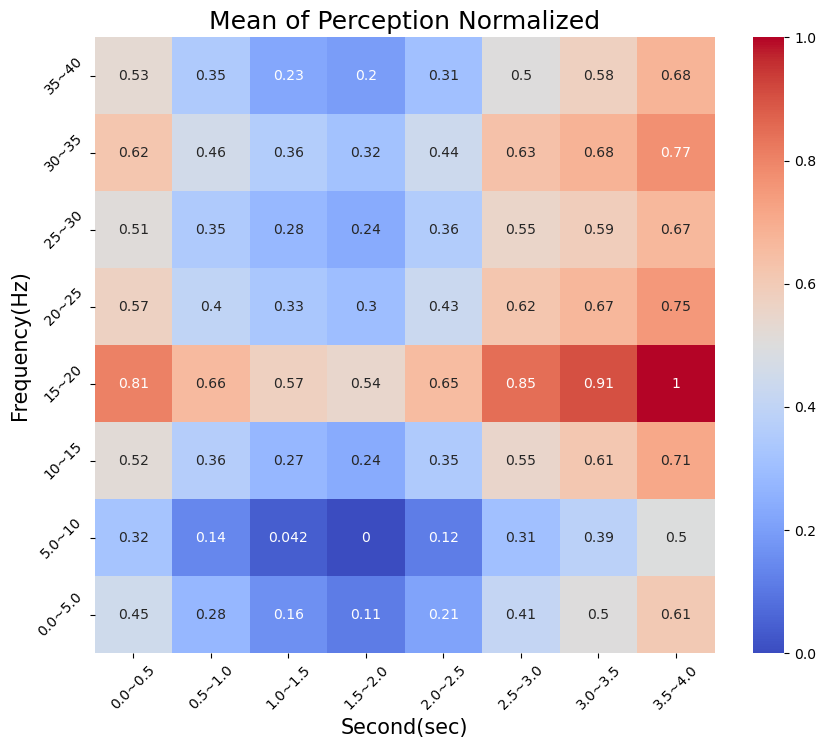}
        \caption{Attention map only for Perception tasks from pre-trained ViT} % Correct the equation
        \label{fig:c}
    \end{subfigure}
    \hfill % Add separation between subfigures
    \begin{subfigure}{.49\textwidth}
        \centering
        \includegraphics[width=\textwidth]{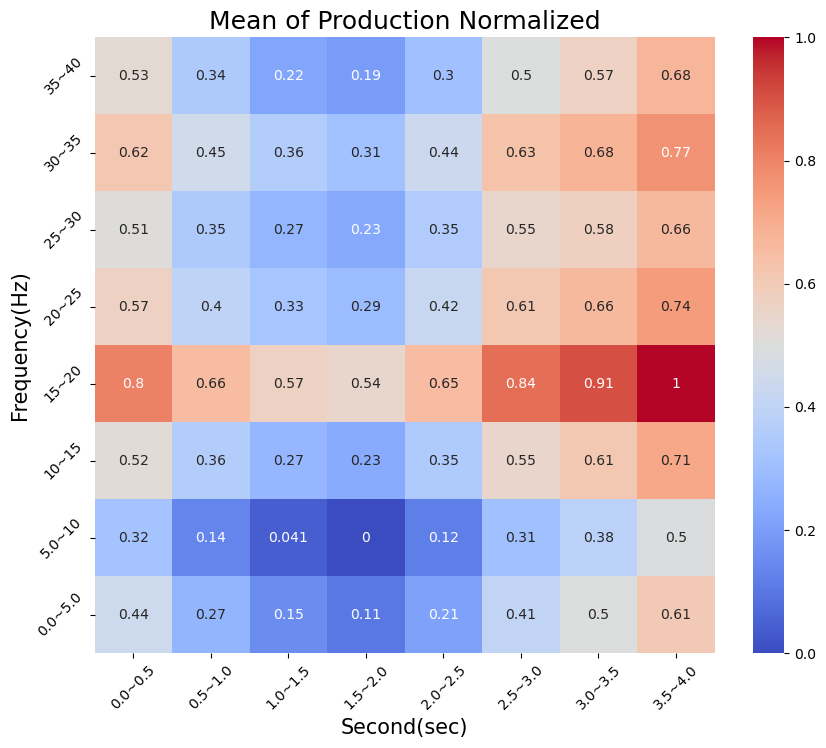}
        \caption{Attention map only for Production tasks from pre-trained ViT} % Correct the equation
        \label{fig:d}
    \end{subfigure}
    
    \caption{Attention Maps of the models during classification. Lower values (indicated by blue) represent regions where the models allocate less attention, whereas higher values, indicated by red, signify areas of focused attention. The x-axis represents the time series from 0 to 4 seconds, and the y-axis represents the frequency series from 0 to 40 Hz.
    Normalized attention maps, averaged from data collected during (a) the perception task (listening) using the Custom ViT, (b) the production task (speaking) using the Custom ViT, (c) the perception task using the pre-trained ViT, and (d) the production task using the pre-trained ViT.}
    \label{fig:three}
\end{figure*}

\begin{figure}[htbp]
    \centering
    \begin{subfigure}[t]{0.49\textwidth}
        \centering
        \includegraphics[width=\textwidth]{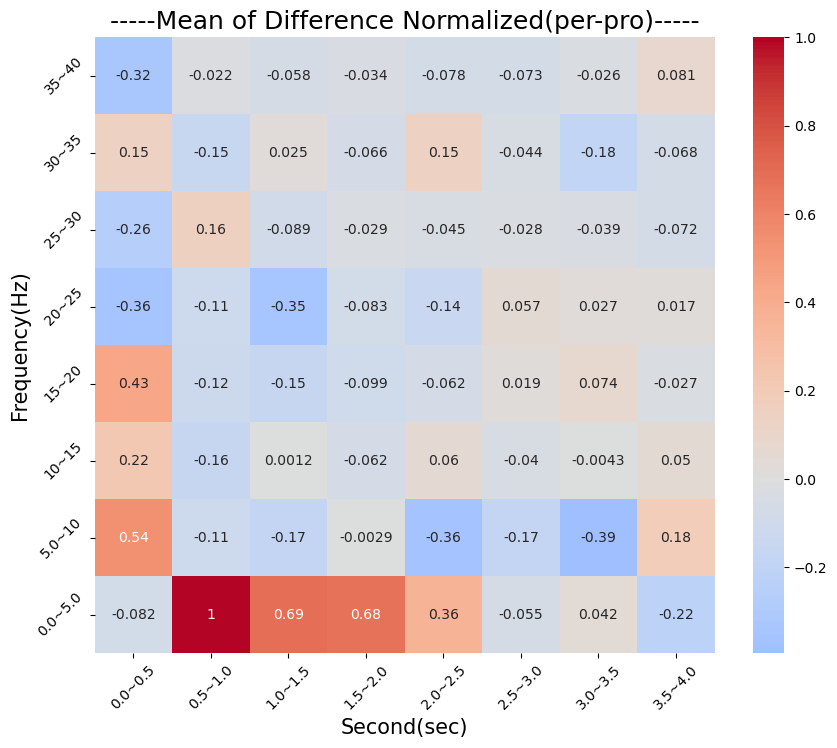}
        \caption{Attention map for Whole tasks from Custom ViT}
        \label{fig:e}
    \end{subfigure}
    \hfill % 図の間にスペースを追加
    \begin{subfigure}[t]{0.49\textwidth}
        \centering
        \includegraphics[width=\textwidth]{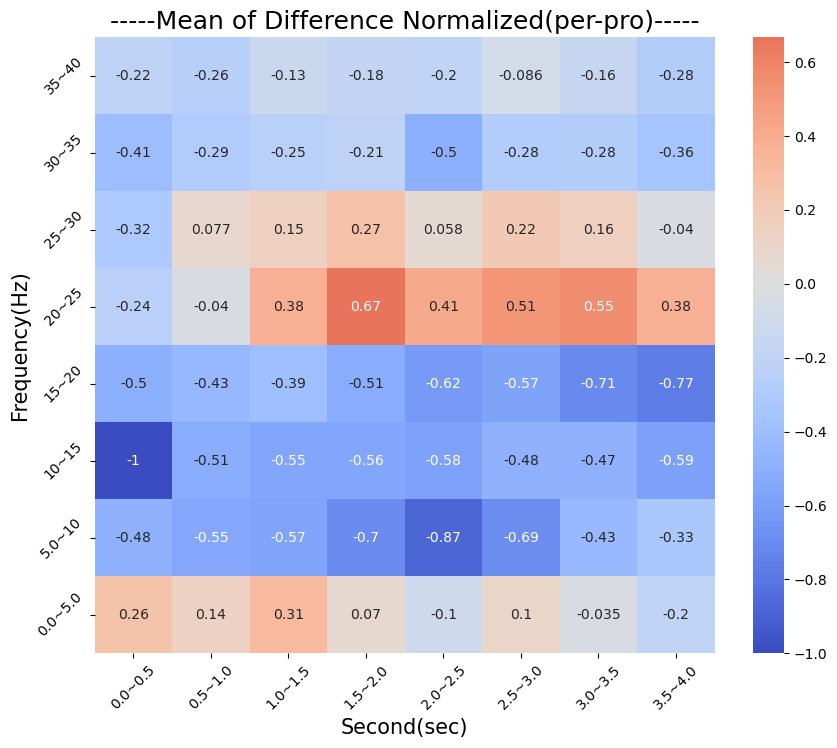}
        \caption{Attention map for Whole tasks from pre-trained ViT}
        \label{fig:f}
    \end{subfigure}
    \caption{Contrasts of attention maps between production and perception tasks. Lower values (blue) indicate greater attention during production tasks, whereas higher values (red) highlight areas of intensified focus during perception tasks. Both axes are consistent with those in Figure 3. Normalized attention maps are obtained by calculating the difference between the Perception and Production attention maps from (a) the Custom ViT and (b) the pre-trained ViT.}
    \label{fig:three graphs}
\end{figure}

Figure \ref{fig:three} highlights that ViTs predominantly focus on the initial stages of the task. The Custom ViT, shown in Figures \ref{fig:a} and \ref{fig:b}, consistently emphasizes the delta and theta bands throughout the task duration, reflecting its sensitivity to lower frequency ranges. In contrast, the pre-trained ViT, depicted in Figures \ref{fig:c} and \ref{fig:d}, exhibits a marked preference for beta waves, particularly the high beta frequencies, and additionally shifts its attention significantly towards the task's conclusion.

Further nuances in the attention distribution are evident from the comparative analyses presented in Figure \ref{fig:three graphs}. The Custom ViT shows the greatest variance between perception and production tasks within the delta band: 0.5 to 4.0Hz, with perceptual tasks showing increased activity in the beta band: 16.5 to 20.0Hz, low beta band: 12.5 to 16Hz, alpha band (8.0 to 12.0Hz), and theta band: 4.0 to 7.0Hz, while production tasks predominantly engage the high beta band: 20.5 to 28Hz and gamma band: over 30Hz. This indicates a complex interplay of frequency bands depending on the cognitive demands of the task, as illustrated in Figure \ref{fig:e}.

Conversely, the pre-trained ViT demonstrates the largest disparities in the alpha band when contrasting perception and production tasks. During perceptual tasks, there is a notable increase in beta band activity, whereas production tasks see heightened activity in the theta, alpha, low beta, and gamma bands. These findings, presented in Figure \ref{fig:f}, suggest the differential engagement of task-dependent brain rhythms, highlighting the adaptability of neural network models to varying cognitive requirements.

These activity patterns underscore the intricate relationship between task-specific cognitive processes and neural focus, as represented by frequency band engagement. The attention maps, particularly those derived from the Custom ViT and pre-trained ViT, validate the hypothesis that neural networks can adaptively highlight relevant EEG features that signify distinct cognitive states associated with specific tasks.
\subsection{Grad-CAM}
\begin{figure}[t!]
\centering
\includegraphics[width=.8\textwidth]{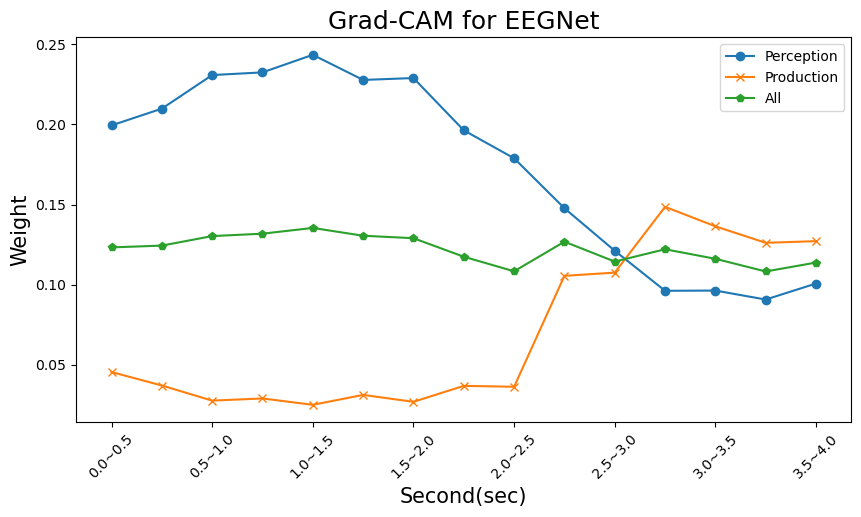}
\caption{The features extracted from the final layer of EEGNet using Grad-CAM.}
\label{fig:gradcam}
\end{figure}

The attention maps derived from EEGNet via Grad-CAM analysis, as shown in Figure \ref{fig:gradcam}, reveal distinct patterns of focus depending on the task and timing. Specifically, when analyzing EEG data associated with perception tasks, the model predominantly concentrates on the initial phase of the task. However, a dominant shift in attention occurs between 2.5 and 3.0 seconds, indicating a temporal transition in neural engagement. This shift suggests that the model identifies critical periods of neural activity that correspond to key moments in the cognitive process, highlighting the dynamic nature of brain function during these tasks.

\subsection{Validity of Usage of ViT for Feature Estimation}

\begin{figure}[t!]
\centering
\includegraphics[width=0.8\textwidth]{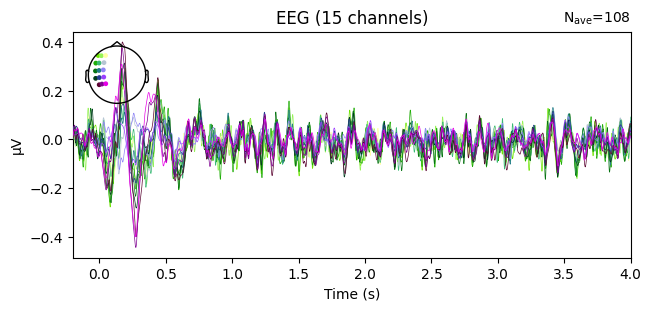}
\caption{Average EEG data across all frequency bands: 1 to 40 Hz, highlighting differences between perception and production tasks. Each trial is aligned to start at 0 seconds, with 0.2 seconds before and 4 seconds after the task onset.}
\label{fig:all}
\end{figure}

\begin{figure}[t!]
\centering
\includegraphics[width=.8\textwidth]{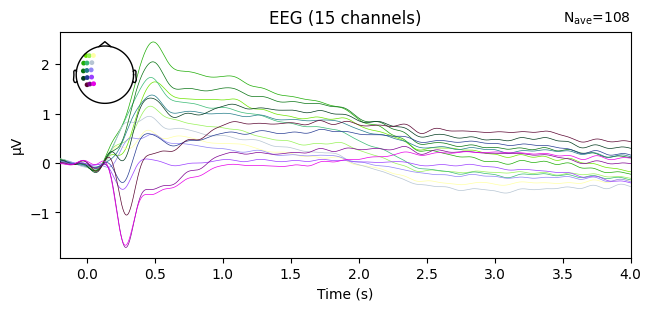}
\caption{Average EEG data within the low-frequency band (1-5 Hz) highlighting differences between perception and production tasks. Each trial is aligned to start at 0 seconds, with 0.2 seconds before and 4 seconds after the task onset.}
\label{fig:diff}
\end{figure}

We further evaluated the EEG data from the previous Grad-CAM and ViT analyses by visualizing the differences between perception and production tasks. This was done by subtracting the EEG signals during the production phase from those during the perception phase, which is the object of classification.

Figure~\ref{fig:all} shows the results of bandpass filtering the EEG data between 1-40 Hz, capturing the broad range of cognitive and neural processes. Each trial is aligned to start at 0 seconds, with an average of 0.2 seconds before and 4 seconds after the task onset across all participants. Significant amplitude differences are observed up to 0.5 seconds after the task starts, indicating early task-specific neural engagement as shown in Figure~\ref{fig:f} and Figure~\ref{fig:gradcam}.

To focus on specific frequency bands and validate the attention map's findings, EEG data from 1 to 5 Hz, including the delta and theta bands, were visualized in Figure~\ref{fig:three graphs} and Figure~\ref{fig:diff}. Each trial is similarly aligned to start at 0 seconds, with 0.2 seconds before and 4 seconds after the task onset averaged across all participants. Differences between perception and production tasks are evident throughout the task duration, underscoring the distinct neural dynamics associated with these cognitive processes.

One potential reason for the early differences observed is the event-related potential (ERP), which captures time-locked neural responses to specific sensory, cognitive, or motor events~\cite{luck2014introduction}. ERPs provide a precise temporal measure of brain activity, which can be effectively captured by EEG. In our study, the ViT model successfully identifies these ERPs, highlighting their significance in distinguishing between perception and production tasks. These visualizations confirm that the attention maps accurately highlight the task-specific neural activity patterns, reinforcing the models' ability to distinguish between different cognitive states based on EEG data.

\section{Discussion}
Our findings underscore the efficacy of neural networks, particularly ViTs, in interpreting EEG features, demonstrating their capacity to recognize EEG characteristics for listening and speaking tasks. Neural network models, especially those equipped with attentional mechanisms, excel at extracting and visualizing salient features from distinct brain activities. This proficiency is derived from their capability~\cite{martinez2021computation} to systematically process inputs across both time and frequency dimensions, thereby preserving the structural integrity of the data.

ViTs demonstrate a unique ability to identify specific features of EEG signals that vary with the cognitive task, enabling precise dissection of frequency and temporal information(\ref{con:c1}). This highlights the adaptability and accuracy of ViTs in neuroscientific research, making them invaluable for tasks requiring a nuanced understanding of brain functions. Our comprehensive evaluation of auditory information in EEG using ViT (\ref{con:c2}) further emphasizes its robustness and effectiveness in capturing task-specific neural dynamics.

However, the performance of each model varies slightly, a phenomenon primarily attributed to the large datasets utilized and the inherent EEG variability among individuals~\cite{thakor2012eeg,chowdhury2023enhancing}. This variability complicates the generalization of findings across different populations and emphasizes the need for models to accommodate individual differences. These observations suggest ample opportunities for further advancements in neural network architectures, potentially enhancing their effectiveness and precision in analyzing complex biological signals.

Interestingly, our study found specific attention areas consistent with previous studies~\cite{giglio2024diverging,kubetschek2021delta,kuhlen2012content,lopez2022state,lin2015eeg}. The synchronization of brain activity, the emergence of features at the start and end of tasks, and the distinctions in frequency bands were clearly illustrated by the attention maps derived from ViTs and EEGNet. These attention maps highlight critical evaluative points for task classification, focusing on distinct EEG features relevant to different cognitive states.

When results were not normalized, performance differences between the Vision Transformers for different tasks may be attributed to the model's insufficient training and the coarse granularity of the training data. While the models aimed to estimate emergent features throughout the tasks, a more detailed, channel-by-channel analysis could potentially improve accuracy.

This study confirms that neural networks can effectively leverage model weights to pinpoint specific EEG features, thereby distinguishing between different cognitive tasks (\ref{con:c1}). This capability validates neural networks' potential to parse EEG data accurately and opens avenues for discovering new insights into EEG features. Such advancements underscore the potential of neural networks to deepen our understanding of the neural bases of cognitive tasks through sophisticated pattern recognition and feature extraction methods.

Moreover, our research suggests that a data-driven approach can reveal how EEG reflects underlying brain activity characteristics (\ref{con:c2}). By analyzing attention maps and model weights, we can infer which aspects of neural activity are most informative for different cognitive states. This approach could lead to identifying biomarkers for specific mental processes, enhancing EEG's diagnostic and therapeutic capabilities in clinical settings.

These findings could significantly improve brain-computer interfaces (BCIs), neurofeedback systems, and other EEG-based diagnostic tools in practical medical applications. For instance, more accurate and individualized EEG analysis could lead to better detection and monitoring of neurological conditions such as epilepsy, depression, and sleep disorders. By providing a clearer understanding of the neural dynamics associated with different tasks, our study paves the way for developing more targeted and effective interventions in cognitive and neurological health.

\section{Conclusion}
This study highlights the potential of neural networks, specifically ViTs and EEGNet, in EEG data interpretation for cognitive task classification. These models recognize established EEG features and uncover new information crucial for understanding brain function. Both Custom ViT and pre-trained ViT demonstrate proficiency in focusing on specific temporal stages of cognitive tasks, with attention to different frequency bands (\ref{con:c2}). EEGNet, analyzed through Grad-CAM, reveals variable attention allocation depending on the task, indicating the temporal complexity involved in processing different cognitive activities. 

The attention maps generated across models are instrumental in understanding how neural networks prioritize certain features for task classification. They identify the EEG signal regions most relevant for distinguishing between cognitive states (\ref{con:c1}). Our study validates the capability of these models in EEG data analysis and suggests that a data-driven approach can reveal significant insights into brain activity patterns (\ref{con:c2}). This paves the way for further enhancements in neural network designs to accommodate individual variability and generalize findings across diverse populations.

In conclusion, using neural networks in EEG analysis offers a transformative approach to understanding the neural bases of cognitive tasks, providing deep insights into the temporal and frequency-related dynamics of brain activity. This research holds promise for improving diagnostic and therapeutic applications in clinical settings, potentially leading to better brain-computer interfaces and neurofeedback systems.

% \section{Acknowledgement}
% Anonymized for the review process.

\bibliographystyle{splncs04}
\bibliography{main}
\end{document}